\documentstyle[12pt]{article}
\begin{document}
\parskip=4pt plus 1pt minus 1pt

\vspace{0.4cm}

\title{
\Large\bf Effect of  Hairpin Diagram on Two-body Nonleptonic $B$
Decays and $CP$ Violation
}

\vspace{0.5cm}
\author{
{\bf Dong-sheng Du}\\
{\normalsize ICTP, P.O. Box 586, 34100 Trieste, Italy}\\
{\normalsize and}\\
\normalsize Institute of High Energy Physics, P.O. Box 918(4), Beijing
100039, P.R. China\thanks{Permanent address}\and
{\bf Zhi-zhong Xing}\\
{\normalsize Institute of High Energy Physics, P.O. Box 918(4), Beijing
100039,  P.R.China}
}
\date{}
\maketitle

\begin{abstract}

	A careful quark-diagram analysis shows that a number of two-body
nonleptonic $B$ decays can occur through the so-called hairpin diagram,
a QCD loop-induced graph different from  penguin  in final-state hadronization
of valence quarks. Using the two-loop renormalization-group-improved
effective Hamiltonian and the naive factorization approximation, we demonstrate
the effect of the hairpin diagram on decay rates and $CP$ asymmetries
for a few interesting channels such as $B^{0}_{d},\bar{B}^{0}_{d}
\rightarrow \psi K_{S}$ and $B^{\pm}_{u}\rightarrow \phi K^{\pm}$.
Branching ratios of some pure hairpin decay modes, e.g.,
$B^{-}_{u}\rightarrow \phi \pi^{-},\phi \rho^{-}$ and
$\bar{B}^{0}_{d}\rightarrow \phi \pi^{0},\phi \rho^{0}, \phi \omega,
\phi \eta,$ etc., are estimated to be on the order of $10^{-7}$.
\end{abstract}

\newpage

\begin{flushleft}
{\Large\bf I $~$ Introduction}
\end{flushleft}

	The study of nonleptonic weak decays of $B$ mesons is very
useful for determining the quark mixing parameters,
probing the origin of $CP$ violation, and investigating the
nonperturbative final-state interactions. In practice, the exclusive
two-body mesonic $B$ decay modes are of more immediate interest
in experiments and can be predicted using approximate or empirical
methods in theory [1]. To the lowest order of the weak interactions, the
physical picture of a $B$ meson decaying into two light mesons may be
described very well by a few distinct quark diagrams [2], and the decay
amplitude can be calculated by using the effective low-energy Hamiltonian
for $\Delta B=\pm 1$ transitions. In employing the recently-presented
two-loop renormalization-group-improved effective Hamiltonian
${\cal H}_{eff}(\Delta B=\pm 1)$ [3] to analyze amplitudes for some
two-body decays of $B$ mesons, we follow the spirit of the spectator
approximation (i.e., neglect Okubo-Zweig-Iizuka (OZI) suppressed
and annihilation terms) and find that there appears a new
term which can not correspond to any of the conventional six quark
diagrams classified by Chau [2,4]. A detailed topological reanalysis shows
that this new term can be described by a color-matched loop diagram,
the so-called hairpin diagram (see Fig.1(d) and Fig.2).
In this letter, we are going to
demonstrate the role of the hairpin diagram in a number of two-body
mesonic $B$ decays. The case for three-body decays will be discussed
elsewhere. By means of the approaches of the quark
diagram analysis and the naive factorization approximation [5], we find
a few pure hairpin channels such as $B^{-}_{u}\rightarrow \phi \pi^{-},
\phi \rho^{-}$ and $\bar{B}^{0}_{d}\rightarrow \phi \pi^{0}, \phi \rho^{0}$,
whose branching ratios are  on the order of $10^{-7}$. We also demonstrate
the hairpin contributions to $\bar{B}^{0}_{d}\rightarrow \psi \bar{K}^{0}$,
$B^{-}_{u}\rightarrow \phi K^{-}$, etc., which are interesting for probing
$CP$ violation in the $B$-meson system [6]. With the help of the future
experimental data on nonleptonic $B$ decays, we are sure that the
detailed analysis made here will be helpful for investigating
the rare nonleptonic $B$ decays, testing the short-distance
QCD calculations at the low-energy scale, and extracting
the information of $CP$ violation.\\

\begin{flushleft}
{\Large\bf II $~$ Quark diagram analysis}
\end{flushleft}

        According to the topology of the lowest order weak interactions
with all QCD strong interaction effects included, the inclusive
nonleptonic decays of a heavy meson can be generally described by
six distinct quark diagrams as illustrated in Ref.[2]. For exclusive
two-body mesonic decays one needs to combine final-state quarks
into specific hadron states. In some graphs an extra quark-antiquark
pair needs to be created in order to produce a pair of light mesons
in the final state. Assuming that $u\bar{u}$, $d\bar{d}$ and
$s\bar{s}$ pairs are produced with equal probability from the
vacuum, as implied by $SU(3)$ symmetry, the final-state quarks
are then allowed to hadronize and arrange themselves in all
possible ways. As a whole, ten distinct quark diagrams can
be formed and contribute
to two-body mesonic $B$ ($D$) decay modes, in which six of them
are either OZI forbidden or formfactor suppressed. Following the
spirit of the spectator model and neglecting those OZI suppressed
and annihilation topologies, we find that there remains four
quark diagrams, as depicted in Fig.1, that may contribute to
a two-body mesonic mode. It is worthwhile to remark that
Fig.1(d), the so-called hairpin diagram, has been ignored in
the previous topological classification [2,4].
In calculations of decay amplitudes by using the traditional
QCD-noncorrected penguin Hamiltonian [4,7], such hairpin contributions
do not appear. In the next section we shall analyze the effect
of the hairpin diagram on some two-body nonleptonic $B$ decays
by means of the two-loop renormalization-group-improved
effective Hamiltonian for $\Delta B=\pm 1$
transitions [3] and the naive factorization approximation [5].

        Now we classify a number of two-body mesonic $B$ decays
which can get contributions from the hairpin diagram. Using the
valence-quark notation the two-body hairpin transition
\begin{equation}
B(b\bar{q}_{s})\longrightarrow X(q_{v}\bar{q}_{v}) + Y(q\bar{q}_{s})
\end{equation}
is illustrated in Fig.2 where $q_{s}=u$ or $d$ denotes the spectator
quark; $q_{v}=u, d, s$ or $c$ denotes the quark from the vacuum; and
$q=d$ or $s$ is from the flavor changing transition $b\rightarrow q$.
We take the wave functions for some $SU(3)$ mesons as follows [8,4]:
\begin{eqnarray}
|\pi^{0}>\; =\; \frac{1}{\sqrt{2}}|\bar{u}u-\bar{d}d>\; ,
& \;\;\; & |\omega>\; =\; \frac{1}{\sqrt{2}}|\bar{u}u+
\bar{d}d>\; ,\nonumber\\
|\rho^{0}>\; =\; \frac{1}{\sqrt{2}}|\bar{u}u-\bar{d}d>\; ,
& \;\;\; & |\eta>\; =\; \frac{1}{\sqrt{3}}|\bar{u}u+
\bar{d}d-\bar{s}s>\; , \\
|a^{0}_{1}>\; =\; \frac{1}{\sqrt{2}}|\bar{u}u-\bar{d}d>\; ,
& \;\;\; & |\eta^{'}>\; =\; \frac{1}{\sqrt{6}}|\bar{u}u+
\bar{d}d+2\bar{s}s>\; ,\nonumber
\end{eqnarray}
and $|\phi>=|\bar{s}s>$. According to the four spectator
quark diagrams illustrated in Fig.1, we
find four groups of two-body mesonic channels in which
the hairpin diagram is involved.

        (i) $~$ Channels occurring only through the hairpin
diagram (Fig.1(d)):
\begin{eqnarray}
B^{-}_{u} & \longrightarrow & \phi \pi^{-}\; ,\; \phi \rho^{-}\; ,\;
\phi a^{-}_{1}\; ; \nonumber\\
\bar{B}^{0}_{d} & \longrightarrow & \phi \pi^{0}\; ,\; \phi \rho^{0}\; ,\;
\phi a^{0}_{1}\; ,\; \phi \omega\; ,\; \phi \eta\; ,
\; \phi \eta^{'}\; .
\end{eqnarray}
It should be noted that
these nine rare decay modes have not been discussed before [1].

        (ii) $~$ Channels occurring through both the penguin diagram
(Fig.1(c)) and the hairpin diagram (Fig.1(d)):
\begin{eqnarray}
B^{-}_{u} & \longrightarrow & \phi K^{-}\; ,\; \phi K^{*-}\; ; \nonumber\\
\bar{B}^{0}_{d} & \longrightarrow & \phi \bar{K}^{0}\; ,\;
\phi \bar{K}^{*0}\; .
\end{eqnarray}
In Ref. [9], these four channels were regarded as the pure
penguin channels and could be used to probe $CP$ violation in
the decay amplitude. We shall see later on that the hairpin
contribution to them is significant in the large $N_{c}$
approximation.

	(iii) $~$ Channels occurring through both the color-mismatched
tree-level diagram (Fig.1(b)) and the hairpin diagram (Fig.1(d)):
\begin{eqnarray}
B^{-}_{u} & \longrightarrow & \psi \pi^{-}\; ,\; \psi \rho^{-}\; ,\;
\psi a^{-}_{1}\; ,\; \psi K^{-}\; ,\; \psi K^{*-}\; ; \nonumber\\
\bar{B}^{0}_{d} & \longrightarrow & \psi \pi^{0}\; ,\; \psi \rho^{0}\; ,\;
\psi a^{0}_{1}\; ,\; \psi \bar{K}^{0}\; ,\; \psi \bar{K}^{*0}\; ,\;
\psi \omega\; ,\; \psi \eta\; ,\; \psi \eta^{'}\; .
\end{eqnarray}
Some of these decay modes, e.g., $\bar{B}^{0}_{d}\rightarrow
\psi \bar{K}^{0}$,  are worthwhile to look at becuase
$B^{0}_{d}$ versus $\bar{B}^{0}_{d}\rightarrow \psi K_{S}$
are very interesting for extracting $CP$-violating signals
in the $B$ system.

	(iv) $~$ Channels occurring through the tree-level diagram(s)
(Fig.1(a) or Fig.1(b)), the penguin diagram (Fig.1(c)) and the
hairpin diagram (Fig.1(d)):
\begin{eqnarray}
B^{-}_{u} & \longrightarrow & \omega \pi^{-}\; ,\; \eta \pi^{-}\; ,\;
\eta^{'} \pi^{-}\; ,\; \omega \rho^{-}\; ,\; \eta \rho^{-}\; ,\;
\eta^{'} \rho^{-}\; ,\; \omega a^{-}_{1}\; ,\; \eta a^{-}_{1}\; ,\;
\eta^{'} a^{-}_{1}\; , \nonumber\\
	  &		& \omega K^{-}\; ,\; \eta K^{-}\; ,\;
\eta^{'} K^{-}\; ,\; \omega K^{*-}\; ,\; \eta K^{*-}\; ,\;
\eta^{'} K^{*-}\; ; \nonumber\\
\bar{B}^{0}_{d} & \longrightarrow & \omega \pi^{0}\; ,\; \eta \pi^{0}\; ,\;
\eta^{'} \pi^{0}\; ,\; \omega \rho^{0}\; ,\; \eta \rho^{0}\; ,\;
\eta^{'} \rho^{0}\; ,\; \omega a^{0}_{1}\; ,\; \eta a^{0}_{1}\; ,\;
\eta^{'} a^{0}_{1}\; , \\
	  &		& \omega \bar{K}^{0}\; ,\; \eta \bar{K}^{0}\; ,\;
\eta^{'} \bar{K}^{0}\; ,\; \omega \bar{K}^{*0}\; ,\; \eta \bar{K}^{*0}\; ,\;
\eta^{'} \bar{K}^{*0}\; , \nonumber\\
          &             & \omega \omega\; ,\; \omega \eta\; ,\;
\omega \eta^{'}\; ,\; \eta \eta\; ,\; \eta \eta^{'}\; ,\;
\eta^{'} \eta^{'}\; .\nonumber
\end{eqnarray}
For our purpose these modes are not as interesting as the
above three groups of modes, since it is difficult to
distinguish the hairpin contribution from the penguin and
tree-level ones in them.\\

\begin{flushleft}
{\Large\bf III $~$ Factorization approximation}
\end{flushleft}

	In this section we use the two-loop low-energy effective
Hamiltonian ${\cal H}_{eff}(\Delta B=\pm 1)$ and the factorization
approach to estimate the contribution of the hairpin diagram to
some of the aforelisted $B$ decay modes. At the physical scale
$\mu =O(m_{b})$, ${\cal H}_{eff}(\Delta B=-1)$ is given by [3]
\begin{equation}
{\cal H}_{eff}(\Delta B=-1)\;=\;\frac{G_{F}}{\sqrt{2}}
\left [V_{ub}V^{*}_{uq}\left (\sum_{i=1}^{2}c_{i}Q^{u}_{i}\right )
+V_{cb}V^{*}_{cq}\left (\sum_{i=1}^{2}c_{i}Q_{i}^{c}\right )
-V_{tb}V^{*}_{tq}\left (\sum_{i=3}^{6}c_{i}Q_{i}\right )\right ]\; ,
\end{equation}
where $V_{jb}V^{*}_{jq}$ ($j=u,c,t;\; q=d,s$) are the
Cabibbo-Kobayashi-Maskawa
(CKM) factors corresponding to $b\rightarrow q$ transition;
$c_{i}$ ($i=1,...,6$) are the QCD-corrected Wilson coefficients at
the physical scale $\mu =O(m_{b})$;
$Q^{u,c}_{i}$ ($i=1,2$) and $Q_{i}$ ($i=3,...,6$) are
the current-current and penguin operators respectively. In the
notation of Ref.[3], we have
\begin{eqnarray}
& Q^{u}_{1} & =\; (\bar{q}_{\alpha}u_{\beta})^{~}_{V-A}
(\bar{u}_{\beta}b_{\alpha})^{~}_{V-A}\; ,\;\;\;\;
Q^{u}_{2}\; =\; (\bar{q}u)^{~}_{V-A}(\bar{u}b)^{~}_{V-A}\; ,\nonumber\\
& Q_{3} & =\; (\bar{q}b)^{~}_{V-A}
\sum_{q^{'}}(\bar{q}^{'}q^{'})^{~}_{V-A}\; ,\;\;\;\;\;
Q_{4}\; =\; (\bar{q}_{\alpha}b_{\beta})^{~}_{V-A}
\sum_{q^{'}}(\bar{q}^{'}_{\beta}q^{'}_{\alpha})^{~}_{V-A}\; ,\\
& Q_{5} & =\; (\bar{q}b)^{~}_{V-A}
\sum_{q^{'}}(\bar{q}^{'}q^{'})^{~}_{V+A}\; ,\;\;\;\;\;
Q_{6}\; =\; (\bar{q}_{\alpha}b_{\beta})^{~}_{V-A}
\sum_{q^{'}}(\bar{q}^{'}_{\beta}q^{'}_{\alpha})^{~}_{V+A}\; ;\nonumber
\end{eqnarray}
and $Q^{c}_{i}$ can be obtained from $Q^{u}_{i}$ through the replacement
$u\rightarrow c$. For the purpose of numerical illustration, we
shall use the following values of $c_{i}$:
\begin{eqnarray}
c_{1}\; =\; -0.324\; , & c_{2}\; =\; 1.151\; , &
c_{3}\; =\; 0.017\; ,\nonumber\\
c_{4}\; =\; -0.038\; , & c_{5}\; =\; 0.011\; , &
c_{6}\; =\; -0.047\; ,
\end{eqnarray}
which are obtained in [3] by taking
$\Lambda^{(4)}_{\overline{MS}}=350$ MeV, $m_{b}=4.8$ GeV and
$m_{t}=150$ GeV.

	Let us estimate branching ratios of a few pure hairpin
channels by taking $B^{-}_{u}\rightarrow \phi \pi^{-}$ for example.
In the naive factorization scheme, we neglect all the
OZI forbidden or annihilation terms and obtain
\begin{eqnarray}
&   & <\phi\pi^{-} |{\cal H}_{eff}(\Delta B=-1)|B^{-}_{u}>\nonumber\\
& = & -\frac{G_{F}}{\sqrt{2}}V_{tb}V^{*}_{td}(a_{4}+a_{6})M^{\phi\pi^{-}}
_{ssd}\; ,
\end{eqnarray}
where $a_{4}$ and $a_{6}$ are the short-distance QCD coefficients
defined by
\begin{equation}
a_{2i-1}\equiv \frac{c_{2i-1}}{N_{c}}+c_{2i}\; ,\;\;\;\;
a_{2i}\equiv c_{2i-1}+\frac{c_{2i}}{N_{c}}
\end{equation}
with $i=1,2,3$; $N_{c}$ is the number of colors; and $M^{\phi\pi^{-}}
_{ssd}$ is the hadronic matrix elements given as
\begin{equation}
M^{\phi\pi^{-}}_{ssd}\equiv <\phi |(\bar{s}s)^{~}_{V-A}|0>
<\pi^{-} |(\bar{d}b)^{~}_{V-A}|B^{-}_{u}>\; .
\end{equation}
One can evaluate $|M^{\phi\pi^{-}}_{ssd}|$ with the help of the
successful empirical approach of Bauer, Stech and Wirbel (BSW) [10].
Considering the approximate rule of discarding $1/N_{c}$ corrections
in exclusive nonleptonic $B$ decays [5], we shall take both $N_{c}=3$
and $N_{c}=\infty$ to give one a feeling of $N_{c}$ dependence
in this factorization approach. Our numerical results of branching
ratios for some pure hairpin channels are listed in Table 1.
We observe that in the case of $N_{c}=3$ both $a_{4}$ and $a_{6}$
are very small so that the values of branching ratios are
vanishingly small. In the case of $N_{c}=\infty$, however,
these branching ratios are on the order of $10^{-7}$.

	The factorized decay amplitudes of $B^{-}_{u}\rightarrow
\phi K^{-}, \phi K^{*-}$ and $\bar{B}^{0}_{d}\rightarrow \phi \bar{K}^{0},
\phi \bar{K}^{*0}$
are  composed of two terms, e.g.,
\begin{eqnarray}
&   & <\phi \bar{K}^{0}|{\cal H}_{eff}(\Delta B=-1)|\bar{B}^{0}_{d}>\nonumber\\
& = & -\frac{G_{F}}{\sqrt{2}}V_{tb}V^{*}_{ts} [a_{3}+(a_{4}+a_{6})]
M^{\phi \bar{K}^{0}}_{sss}\; ,
\end{eqnarray}
which get contributions from the penguin and hairpin diagrams,
respectively. Using the values of $c_{i}$ given in Eq.(9), we obtain
the ratio of the hairpin amplitude to the penguin one for these four
modes:
\begin{equation}
\left |\frac{a_{4}+a_{6}}{a_{3}}\right |\approx \left \{
\begin{array}{ll}
1.2\%  & \;\; N_{c}=3 \\
73.7\% & \;\; N_{c}=\infty\; .
\end{array}\right .
\end{equation}
Obviously, the hairpin contribution to decay amplitudes is significant
if one discards those $1/N_{c}$ terms. Although the above estimate
is based on the naive factorization approximation and can not be
strictly justified, it shows that in these penguin-hairpin mixed rare
decays the latter contributions may be important and should be
considered seriously.

	Conventionally a two-body decay mode of $B^{-}_{u}$ (or
$\bar{B}^{0}_{d}$) into $\psi$ and a $SU(3)$ meson is analyzed at the
tree level [4,7,9]. Employing ${\cal H}_{eff}(\Delta B=-1)$ in Eq.(7),
here we examine
the influence of the hairpin diagram on the overall decay amplitudes
of such channels. Taking $\bar{B}^{0}_{d}\rightarrow
\psi \bar{K}^{0}$ for example, we obtain
\begin{eqnarray}
&   & <\psi \bar{K}^{0} |{\cal H}_{eff}(\Delta
B=-1)|\bar{B}^{0}_{d}>\nonumber\\
& = & \frac{G_{F}}{\sqrt{2}}\left [ V_{cb}V^{*}_{cs}a_{2}
-V_{tb}V^{*}_{ts} (a_{4}+a_{6})\right ]M^{\psi\bar{K}^{0}}_{ccs}\; .
\end{eqnarray}
The ratio of the hairpin amplitude to the tree-level one is
\begin{equation}
\left |\frac{V_{tb}V^{*}_{ts}}{V_{cb}V^{*}_{cs}}\right |\cdot
\left |\frac{a_{4}+a_{6}}{a_{2}}\right | \approx
\left (1+\frac{\lambda^{2}}{2}\right )
\left |\frac{a_{4}+a_{6}}{a_{2}}\right | \approx \left \{
\begin{array}{ll}
0.7\%  & \;\; N_{c}=3 \\
8.8\%  & \;\; N_{c}=\infty\; ,
\end{array}\right .
\end{equation}
where $\lambda =0.22$ is a CKM parameter [11]. We observe that in the case
of $N_{c}=\infty$ the hairpin amplitude is not too small compared
with the tree-level amplitude, and this enhancement merely comes from the
short-distance QCD.

	As for the charmless two-body decay modes listed in Eq.(6),
the charged $B$ channels can occur through all the four quark
diagrams of Fig.1, while the neutral $B$ transitions get contributions
from Fig.1(b-d). Considering the fact that
\begin{equation}
\left |\frac{V_{tb}V^{*}_{td}}{V_{ub}V^{*}_{ud}}\right |\sim 1\; ,\;\;\;\;
\left |\frac{V_{tb}V^{*}_{ts}}{V_{ub}V^{*}_{us}}\right |
\sim \frac{1}{\lambda^{2}}\; ,
\end{equation}
the effect of the penguin and hairpin amplitudes
are expected to be very large for those modes corresponding to
$b\rightarrow (u\bar{u})s$ such as $B^{-}_{u}\rightarrow
\omega K^{-}$ and $\bar{B}^{0}_{d}\rightarrow \eta \bar{K}^{*0}$,
and to be comparable to the tree-level ones in those modes
corresponding to $b\rightarrow (u\bar{u})d$ such as $B^{-}_{u}
\rightarrow
\eta \rho^{-}$ and $\bar{B}^{0}_{d}\rightarrow \omega
a^{0}_{1}$. Since these channels are tree-loop mixed, they are not
very interesting for probing the loop-induced effects in branching
ratios and $CP$ asymmetries.\\

\begin{flushleft}
{\Large\bf IV $~$ Discussion}
\end{flushleft}

	Now we are in a position to discuss the effect of the hairpin
diagram on $CP$ asymmetries for a few interesting $B$ decay modes such
as $B^{0}_{d},\bar{B}^{0}_{d}\rightarrow \psi K_{S}$ and
$\phi K_{S}$. To a high degree of accuracy in the standard model,
the time-integrated partial rate asymmetry of $B^{0}_{\rm phys}$
and $\bar{B}^{0}_{\rm phys}$ decaying into a $CP$ eigenstate $f$
is given by [12]
\begin{equation}
{\cal A}_{CP}\; =\; \frac{(1-|\xi_{f}|^{2})-2x{\rm Im}\left (
\displaystyle\frac{q}{p}\xi_{f}\right )}{(1+x^{2})(1+|\xi_{f}|^{2})}\; ,
\end{equation}
where $x\equiv \Delta m/\Gamma$ is the $B^{0}-\bar{B}^{0}$ mixing
parameter; and
\begin{equation}
\xi_{f}\; \equiv \; \frac{A(\bar{B}^{0}\rightarrow f)}
{A(B^{0}\rightarrow f)}\; ,\;\;\;\; \frac{q}{p}\; =\;
\frac{V^{*}_{tb}V_{td}}{V_{tb}V^{*}_{td}}
\end{equation}
denote the ratio of transition amplitudes and mixing phase, respectively.
If all the contributing components of $A(B^{0}\rightarrow f)$ have
the same weak phase or the same strong phase shift, then $\xi_{f}$
will be a pure weak phase and a nonzero ${\cal A}_{CP}$ will arise
only from $B^{0}-\bar{B}^{0}$ mixing [13].

	In contrast with the previous discussions in which $B^{0}_{d}
\rightarrow \psi K_{S}$ was assumed to occur through only the
tree-level diagram [6,13], our analysis induces the hairpin contribution
to this mode (see Eq.(15)). As a result,
\begin{equation}
\xi_{\psi K_{S}}\; =\; -\frac{V^{*}_{cb}V_{cs}a_{2}-V^{*}_{tb}V_{ts}
(a_{4}+a_{6})}{V_{cb}V^{*}_{cs}a_{2}-V_{tb}V^{*}_{ts}(a_{4}+
a_{6})}\cdot \frac{V^{*}_{cs}V_{cd}}{V_{cs}V^{*}_{cd}} \; ,
\end{equation}
where we have taken into account the $K^{0}-\bar{K}^{0}$ mixing
phase, and the minus sign comes from a $CP$ transformation for
the $CP$-odd state $\psi K_{S}$. In terms of the CKM
parameters [11], we obtain
\begin{equation}
\frac{V_{tb}V^{*}_{ts}}{V_{cb}V^{*}_{cs}}\; \approx\;
\left (1+\frac{\lambda^{2}}{2}\right )(1-i\eta \lambda^{2})\; ,
\;\;\;\; \frac{V_{cs}V^{*}_{cd}}{V_{cb}V^{*}_{cs}}\; \approx \;
-\frac{1}{A\lambda}(1-i\eta \lambda^{2})\; .
\end{equation}
The above equation shows that the relative phase shifts among
$V_{cb}V^{*}_{cs}, V_{tb}V^{*}_{ts}$ and $V_{cs}V^{*}_{cd}$
are $O(\eta\lambda^{2})$ suppressed. Accordingly $\xi_{\psi K_{S}}
\approx 1+iO(\eta\lambda^{2})$, and ${\rm Im}(\frac{q}{p}\xi_{\psi
K_{S}})\approx {\rm Im}(\frac{q}{p})$ remains to a good accuracy.
This means that the effect of the hairpin diagram on the $CP$
asymmetry of $B^{0}_{d}$ versus $\bar{B}^{0}_{d}\rightarrow
\psi K_{S}$ is unimportant, although its influence on the
decay rates of these two modes are not too small in the large
$N_{c}$ approximation. Note that the penguin and hairpin
transitions have different isospin structure from the tree-level
one, thus the above estimates of the hairpin contribution
to decay rates and $CP$ asymmetries can only serve as an illustration.
Anyway $B^{0}_{d},\bar{B}^{0}_{d}\rightarrow \psi K_{S}$
remain the most interesting decay modes for probing $CP$ violation
with little hadronic uncertainties. In general, however, we
can not expect a little influence of the hairpin diagram on $CP$
asymmetries in the charmless nonleptonic $B$ decays listed
in Eq.(6).

   In Ref.[9], the decay modes $B^{\pm}_{u}\rightarrow
\phi K^{\pm}$ were predicted by using the QCD-noncorrected
penguin Hamiltonian and expected to manifest large $CP$
violation in the decay amplitude. Taking into account the
hairpin contribution, here we employ the two-loop
renormalization-group-improved effective Hamiltonian in
estimating decay rates for such charmless channels.
As we have seen in Eq.(7), however, the penguin operators
$Q_{3,...,6}$ are only present in the $u-t$ sector and are
thus all proportional to the same CKM factor $V_{tb}V^{*}_{tq}$.
As a result, the partial decay rate difference of $B^{+}_{u}$ and
$B^{-}_{u}$ vanishes in principle. This puzzle arises from
the form of effective Hamiltonian we take in calculations,
and is of course unreasonable. In Ref.[14], an attempt
was made to generate penguin phases in both the $u-t$ and $u-c$
sectors by inducing the one-loop penguin matrix elements
of $Q_{2}$ operators and removing the renormalization scheme
dependence of the Wilson coefficient functions including
next-to-leading QCD corrections. Here we do not want to
follow such an argument to recover $CP$ asymmetries for
the aforementioned charmless channels, since it is
unjustified and untrustworthy. For our purpose we
want to show that the hairpin contributions are significant
in these rare decay modes and may induce $CP$ violation
in the decay amplitude. A deeper investigation of this
problem will be made elsewhere.\\

\begin{flushleft}
{\Large\bf V $~$ Conclusions}
\end{flushleft}

	We have made a brief quark-diagram analysis to show that,
generally, the QCD loop-induced hairpin diagram can contribute
to two-body nonleptonic $B$ decays( of course, also to multi-body
nonleptonic $B$ decays). In the naive factorization calculations for
decay amplitudes, the hairpin component appears if one employs the
two-loop renormalization-group-improved effective Hamiltonian instead
of the traditional QCD-noncorrected penguin Hamiltonian. We find a number
of exclusive channels occurring only through the hairpin diagrams, whose
branching ratios are estimated to be on the order of $10^{-7}$ in the large
$N_{c}$ approximation. These rare decay modes , if they were measured in the
near future, could be used to test the short-distance QCD calculations
for $\Delta B=\pm 1$ decays. Although the effect of the hairpin
diagram on decay rates for $B^{0}_{d},\bar{B}^{0}_{d}\rightarrow
\psi K_{S}, \phi K_{S}$ and $B^{\pm}_{u}\rightarrow \phi K^{\pm}$
are not too small, its influence on $CP$ asymmetries for these
interesting channels is still expected to be very small in our
rough estimates. At present, the decay modes $B^{0}_{d}$ versus
$\bar{B}^{0}_{d}\rightarrow \psi K_{S}$ remain the cleanest
nonleptonic channel for probing $CP$ violation in the $B$ system,
since their partial rate asymmetry is approximately independent
of hadronic uncertainties. For a number of charmless decay modes,
however, our theoretical techniques for calculating their decay rates
are still problematic, especially in treating hadronic matrix
elements. \\
\begin{flushleft}
{\Large\bf Acknowledgement}
\end{flushleft}

	One of the authors (Du) would like to thank ICTP at Trieste for
its hospitality. This work was also supported in part by the National
Natural Science Foundation of China.

\newpage

\begin{flushleft}
{\Large\bf Figure Captions}
\end{flushleft}

\vspace{0.4cm}

{\bf Fig.1} $~$ Four distinct spectator quark diagrams for two-body
nonleptonic $B$ decays at the scale $\mu =O(M_{W})$:

	$(a)$ the color-matched tree-level diagram;

	$(b)$ the color-mismatched tree-level diagram;

	$(c)$ the penguin diagram;

	$(d)$ the hairpin diagram. \\

{\bf Fig.2} $~$ The hairpin diagram for $B(b\bar{q}_{s})\rightarrow
X(q_{v}\bar{q}_{v})+Y(q\bar{q}_{s})$ transitions at the scale
$\mu =O(m_{b})$.

\newpage

\begin{flushleft}
{\Large\bf Table Caption}
\end{flushleft}

\vspace{0.4cm}

{\bf Table 1} $~$ Branching ratios for a few pure hairpin
decay modes of $B$ mesons. Here the CKM parameters [11] are taken
as $\lambda=0.22$, $A=1.0$, $\rho=-0.4$ and $\eta=0.25$ [15];
and the relevant values of decay constants and formfactors are
quoted from [10].\\

\begin{center}
{\bf Table 1}
\end{center}

\begin{center}
\begin{tabular}{|l|c|c|} \hline\hline
Decay mode & \multicolumn{2}{c|}{ Branching ratio } \\ \cline{2-3}
	   & $N_{c}=3$ & $N_{c}=\infty$ 	\\ \hline
$B^{-}_{u}\rightarrow \phi \pi^{-}$ &
$5.4\times 10^{-11}$ & $3.8\times 10^{-7}$ \\
$B^{-}_{u}\rightarrow \phi \rho^{-}$ &
$7.6\times 10^{-11}$ & $5.3\times 10^{-7}$ \\
$B^{-}_{u}\rightarrow \phi a_{1}^{-}$ &
$6.6\times 10^{-11}$ & $4.6\times 10^{-7}$ \\
$\bar{B}^{0}_{d}\rightarrow \phi \pi^{0}$ &
$2.7\times 10^{-11}$ & $1.9\times 10^{-7}$ \\
$\bar{B}^{0}_{d}\rightarrow \phi \rho^{0}$ &
$3.8\times 10^{-11}$ & $2.7\times 10^{-7}$ \\
$\bar{B}^{0}_{d}\rightarrow \phi a_{1}^{0}$ &
$3.3\times 10^{-11}$ & $2.3\times 10^{-7}$ \\
$\bar{B}^{0}_{d}\rightarrow \phi \omega$ &
$3.7\times 10^{-11}$ & $2.6\times 10^{-7}$ \\
$\bar{B}^{0}_{d}\rightarrow \phi \eta$ &
$1.5\times 10^{-11}$ & $1.0\times 10^{-7}$ \\
$\bar{B}^{0}_{d}\rightarrow \phi \eta^{'}$ &
$1.9\times 10^{-11}$ & $1.3\times 10^{-7}$ \\ \hline\hline
\end{tabular}
\end{center}

\end{document}